\documentclass{iau} 
\usepackage{graphicx}

\title[The origin and impact of Wolf-Rayet-type mass loss] 
{The origin and impact of Wolf-Rayet-type mass loss}

\author[A.A.C. Sander et al.]   
{Andreas A.C. Sander$^1$,
  Jorick S. Vink$^2$,
  Erin R. Higgins$^2$,
  Tomer Shenar$^3$,
  Wolf-Rainer Hamann$^4$,
  and Helge Todt$^4$}

\affiliation{$^1$Zentrum für Astronomie der Universität Heidelberg, Astronomisches Rechen-Institut, Mönchhofstr. 12-14, 69120 Heidelberg, Germany \\ email: {\tt andreas.sander@uni-heidelberg.de} \\[\affilskip]
$^2$Armagh Observatory and Planetarium, College Hill, Armagh BT61 9DG, N.\ Ireland\\[\affilskip]
$^3$Anton Pannekoek Institute for Astronomy, Science Park 904, 1098 XH, Amsterdam, The Netherlands\\[\affilskip]
$^4$Institut für Physik und Astronomie, Universität Potsdam, Karl-Liebknecht-Str. 24/25, D-14476 Potsdam, Germany
}

\pubyear{2022}
\volume{366}  
\jname{The Origin of Outflows in Evolved Stars}
\editors{L.\ Decin, A.\ Zijlstra, \& C.\ Gielen, eds.}
\begin{document}

\maketitle

\begin{abstract}
Classical Wolf-Rayet (WR) stars mark an important stage in the late evolution of massive stars. As hydrogen-poor massive stars, these objects have lost their outer layers, while still losing further mass through strong winds indicated by their prominent emission line spectra. Wolf-Rayet stars have been detected in a variety of different galaxies. Their strong winds are a major ingredient of stellar evolution and population synthesis models. Yet, a coherent theoretical picture of their strong mass-loss is only starting to emerge. In particular, the occurrence of WR stars as a function of metallicity (Z) is still far from being understood.

To uncover the nature of the complex and dense winds of Wolf-Rayet stars, we employ a new generation of model atmospheres including a consistent solution of the wind hydrodynamics in an expanding non-LTE situation. With this technique, we can dissect the ingredients driving the wind and predict the resulting mass-loss for hydrogen-depleted massive stars. Our modelling efforts reveal a complex picture with strong, non-linear dependencies on the luminosity-to-mass ratio and Z with a steep, but not totally abrupt onset for WR-type winds in helium stars. With our findings, we provide a theoretical motivation for a population of helium stars at low Z, which cannot be detected via WR-type spectral features. Our study of massive He-star atmosphere models yields the very first mass-loss recipe derived from first principles in this regime. Implementing our first findings in stellar evolution models, we demonstrate how traditional approaches tend to overpredict WR-type mass loss in the young Universe. 

\end{abstract}

\firstsection 
\section{Introduction}

The striking appearance of an emission-line dominated optical spectrum has led to the introduction of a new spectral class in the 19th century: the Wolf-Rayet (WR) stars. This purely morphological definition has led to a situation that objects of different evolutionary stages can actually fall into this definition. This includes for example a subgroup of central stars of planetary nebulae showing WR-type spectra. In the massive star regime, the WR phenomenon occurs at multiple evolutionary stages. Among those are the now-called \emph{classical} Wolf-Rayet (cWR) stars, which are evolved, core-He burning objects, partly or completely depleted in hydrogen. Another class are less evolved very massive stars that show WR-type spectra of the WNh subclass which are believed to be core-H burning. The latter essentially form an extension of the main sequence \cite[(Crowther \& Walborn 2011)]{CW2001} for very high initial masses, although the spectral WNh-type alone is not sufficient to infer the evolutionary status \cite[(see, e.g., the case of R144 in Shenar et al.\ 2021)]{Shenar+2021}.

Their emission-line spectrum turns out to be a challenge for determining the stellar parameters of WR stars, in particular for the radii and the masses. Contrary to OB-type stars, the surface gravity cannot be determined via the wings of absorption lines. For WR stars at about solar metallicity, the whole spectrum is often formed in the wind, making it impossible to deduce a stellar radius (and a corresponding temperature) from spectral fitting due to inherent degeneracies \cite[(e.g.\ Najarro et al.\ 1997, Hamann \& Gr{\"a}fener 2004)]{Najarro+1997,HamannGraefener+2004}. This degeneracy can be arbitrarily broken by invoking a fixed assumption -- usually a $\beta$-law -- about the velocity field $v(r)$, but often results in discrepancies between empirical radii and expectations from stellar structure models (nowadays also termed the ``WR radius problem''). These obstacles in pinning down important parameters of WR stars also limit the capabilities of empirical descriptions of WR mass loss as the measurements of the mass-loss rate $\dot{M}$ need to be associated with proper stellar parameters, which ideally should agree with predictions from stellar structure and evolution calculations.

\section{Hydrodynamically-consistent atmosphere models}

If the spectrum is completely formed in rapidly expanding layers, the location of any stellar radius can only be inferred indirectly. In principle, this could be done by stellar structure calculations \cite[(e.g.\ Grassitelli et al.\ 2018, Ro 2019, Poniatowski et al.\ 2021)]{Grassitelli+2018,Ro2019,Poniatowski+2021}, but would require a treatment of the moving layers where the hydrostatic equation is no longer valid due to a non-zero inertia term \cite[(see Grassitelli et al.\ 2018 for such an approach)]{Grassitelli+2018}. Moreover, the presence of a non-negligible velocity leads to a difference between the flux-mean opacity and the Rosseland opacity. While the latter is available in tabulated forms \cite[(e.g.\ the widely used OPAL tables by Iglesias \& Rogers 1996)]{IglesiasRogers1996}, such that a detailed calculation can be avoided, the flux-mean opacity in a medium outside of (local) thermodynamical equilibrium requires a frequency-dependent calculation including the necessary determination of the radiation field and the population numbers. 

For these reasons, stellar atmosphere codes provide a promising opportunity. Current codes such as PoWR \cite[(Gr{\"a}fener et al. 2002, Hamann \& Gr{\"a}fener 2003)]{Graefener+2002,HamannGraefener2003} contain the necessary physics for an expanding non-LTE environment, but need to be extended/augmented to solve the hydrodynamic (HD) equation of motion consistently \cite[(Gr{\"a}fener \& Hamann 2005, Sander et al.\ 2017)]{GraefenerHamann2005,Sander+2017}. Studies performed with HD-consistent versions of PoWR have revealed the complex shape of wind velocity fields in classical WR stars \cite[(Gr{\"a}fener \& Hamann 2005, Sander et al.\ 2020, see also Fig.\,\ref{fig:wrtype-mdot})]{GraefenerHamann2005,Sander+2020}, which are a consequence of the various ionization changes in a WR wind. These studies also revealed the fundamental importance of the iron opacities, which are responsible for launching the wind in cWR stars. The fundamental role of iron including its scaling of WR-type mass loss was also found in independent Monte Carlo simulations made by \cite[Vink \& de Koter (2005)]{VinkdeKoter2005}.

\section{Mass-loss rates for classical WN stars}

To investigate the mass-loss of classical WN stars in a more detailed way, we calculated sequences of HD-consistent PoWR models for hydrogen-free stars. Using a fixed stellar temperature of $T_\ast = 141\,$kK, each sequence adopted a different metallicity $Z$, ranging from $2\,Z_\odot$ down to $0.02\,Z_\odot$. To reduce the number of free parameters, we invoked the mass-luminosity relation for hydrogen-free stars from \cite[Gr{\"a}fener et al.\ (2011)]{Graefener+2011}. With these assumptions, our model stars approximately correspond to stars on the theoretical He ZAMS. We calculated models with masses up to $400\,M_\odot$ in order to investigate the asymptotic behavior of $\dot{M}$. This turned out to be crucial for extracting a mathematical description which can be extrapolated more realistically than power-law descriptions. Details of the modeling efforts and parameter ranges are given in \cite[Sander \& Vink (2020)]{SanderVink2020}.

\begin{figure}[htb]
\begin{center}
   \includegraphics[width=0.62\textwidth]{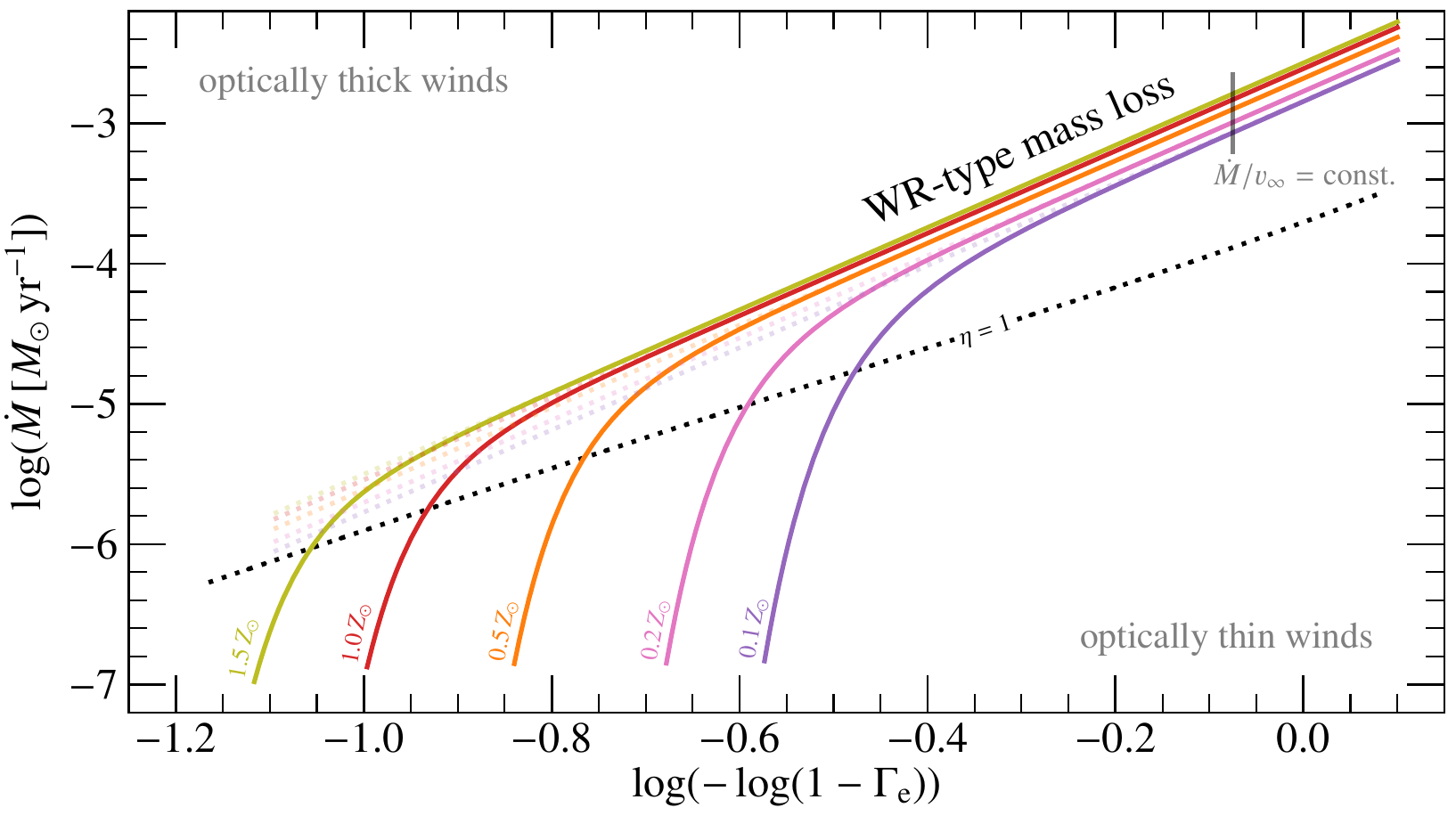} \hfill
 	 \includegraphics[width=0.37\textwidth]{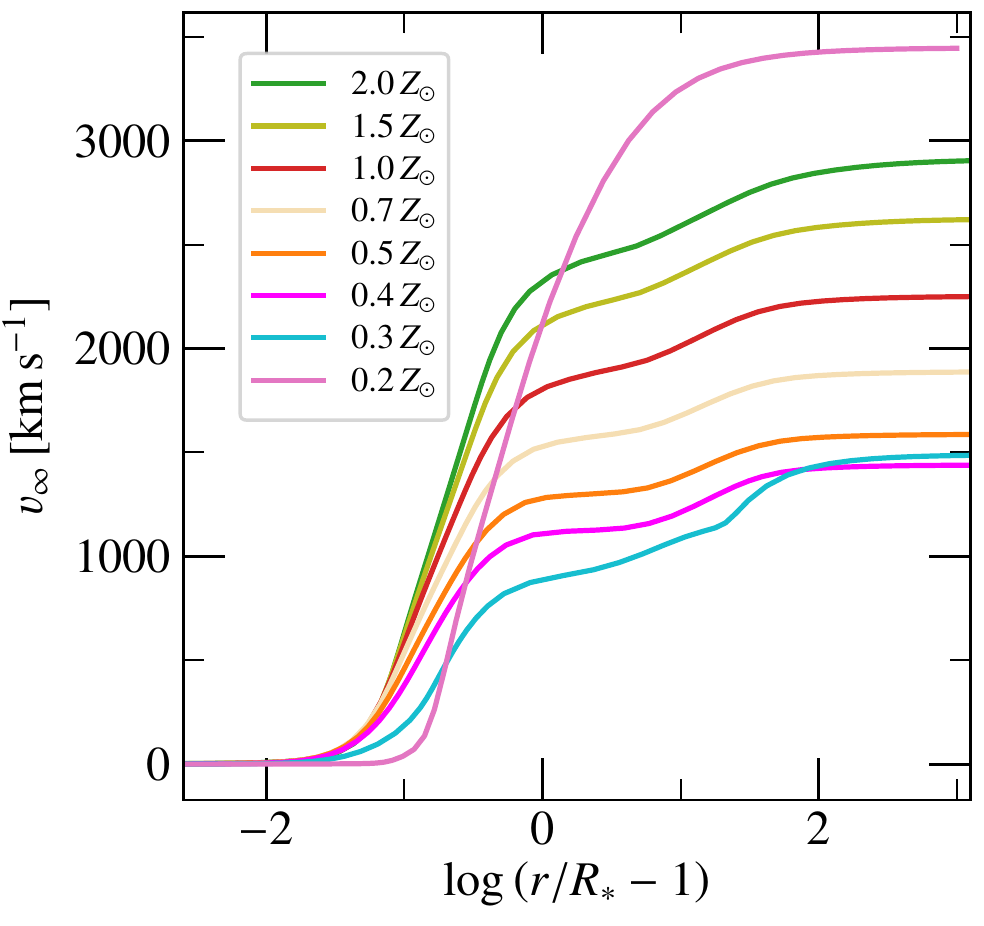}
 \caption{Left panel: Mass-loss $\dot{M}$ rate of He stars for different metallicities based on fitting the results from the hydrodynamic model calculations (full lines). The linear trend in the regime of totally optically thick winds (thin dashed lines) abruptly changes due to the transition to optically thin winds. This coincides with a breakdown of $\dot{M}$. In the optically thick regime, the ratio of $\dot{M}$ over terminal velocity $v_\infty$ is the same for all metallicities. $\Gamma_\mathrm{e} \propto L/M$ denotes the Eddington parameter which is proportional to the ratio of luminosity $L$ over mass $M$.\\[0.2em]
 Right panel: Velocity fields from HD modelling for a $20\,M_\odot$ He star at different metallicities $Z$.}
   \label{fig:wrtype-mdot}
\end{center}
\end{figure}

\begin{figure}[htb]
\begin{center}
   \includegraphics[width=0.57\textwidth]{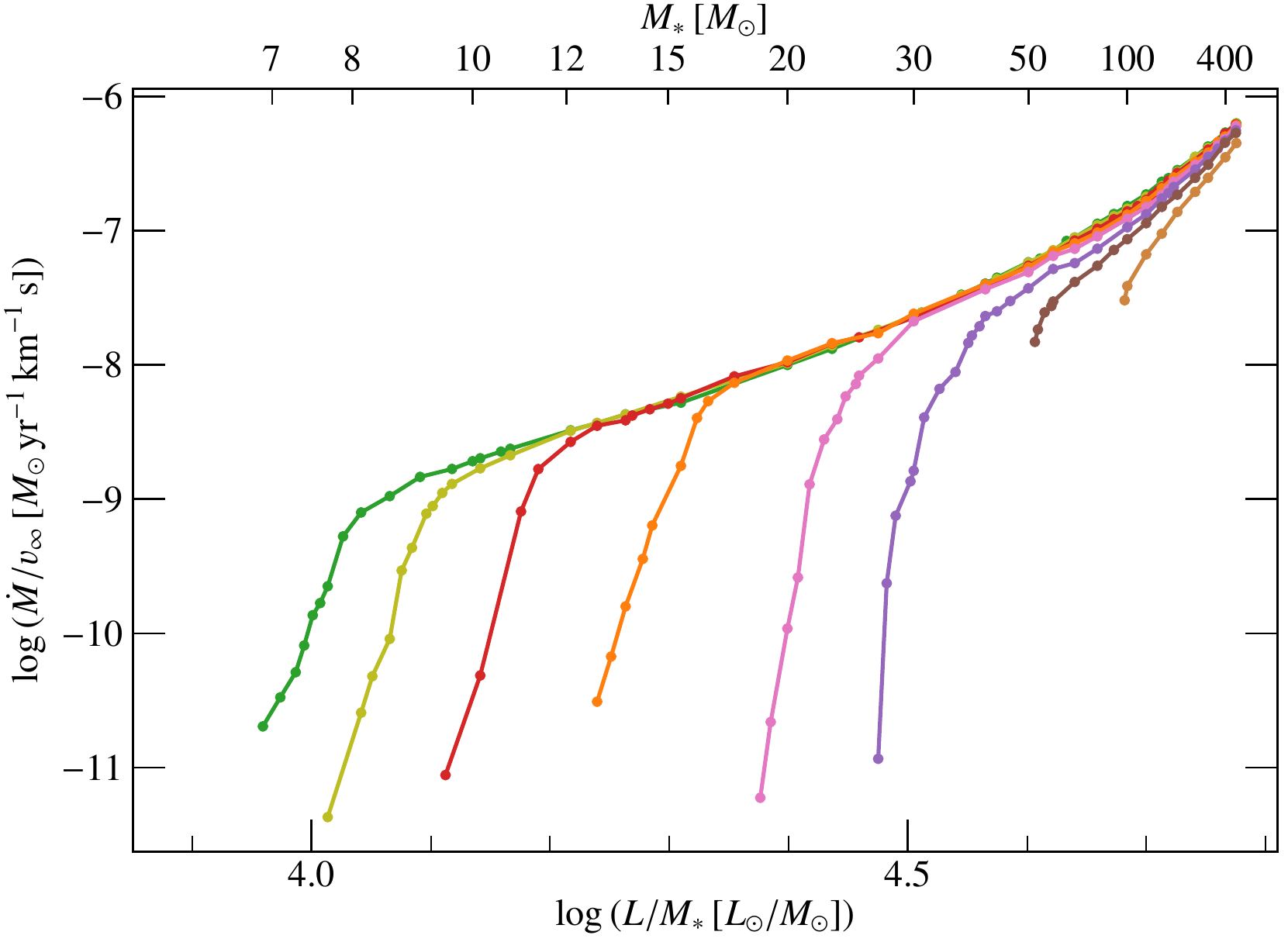} \hfill
	 \includegraphics[width=0.4\textwidth]{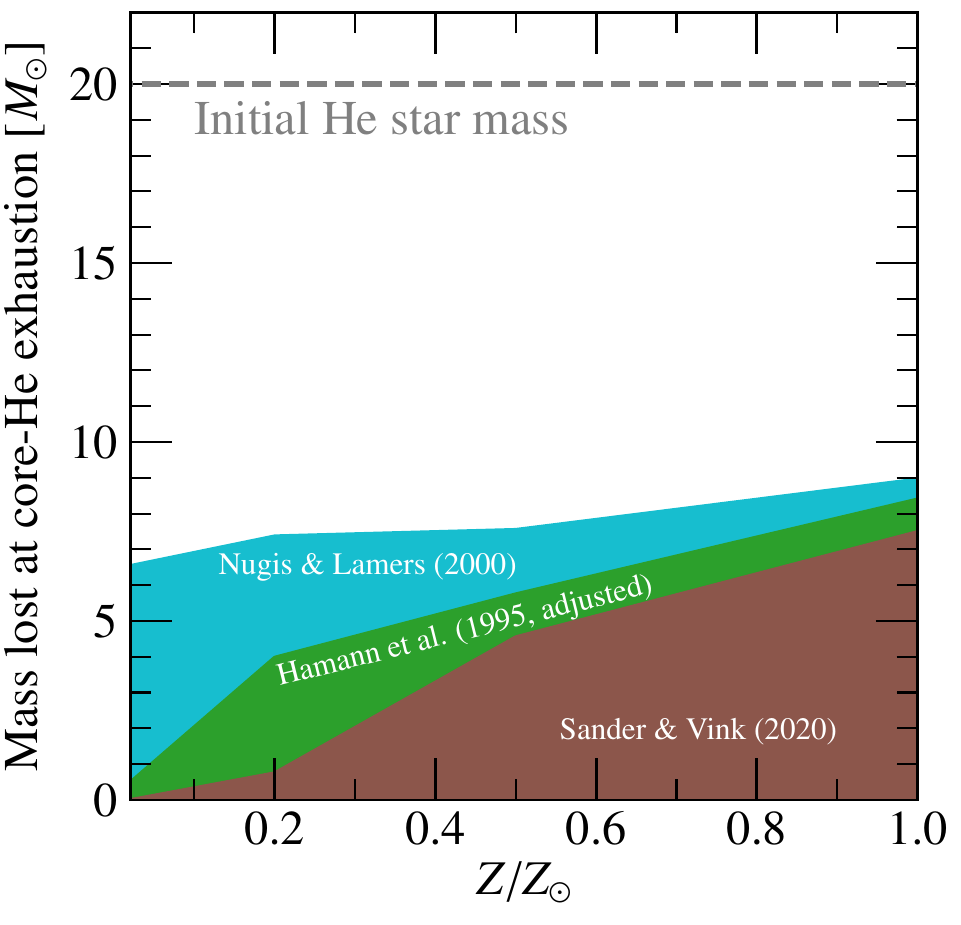} 
\caption{Left panel: Resulting data from dynamically consistent atmosphere models for the ratio of the mass-loss rate $\dot{M}$ over the terminal velocity $v_\infty$ as a function of the luminosity-to-mass ratio $L/M$: The trends for different metallicities (indicated by different colors) align in the limit of optically thick winds where the wind is launched in deep layers (optical depth $\tau \gg 1$).\\[0.1em]
   Right panel: Total mass lost due to stellar winds for a $20\,M_\odot$ He star until core-He exhaustion at different metallicities $Z$ as calculated in the evolutionary models in \cite[Higgins et al.\ (2021)]{Higgins+2021}. The three colored areas reflect the total mass lost according to the three labeled $\dot{M}$ recipes.}
   \label{fig:wrtype-mdotdvinf}
\end{center}
\end{figure}

Our findings are summarized in Fig.\,\ref{fig:wrtype-mdot}, where we plot the resulting curves from fitting our data points from the HD model calculations. In the limit of dense winds, WR-type mass loss can be described by a linear relation between $\log\dot{M}$ and $\log\left[-\log\left(1-\Gamma_\mathrm{e}\right)\right]$ with different metallicities only causing different offsets. Moreover, all these linear curves align when considering $\log(\dot{M}/v_\infty)$ instead of $\log\dot{M}$ (see also Fig.\,\ref{fig:wrtype-mdotdvinf}). This means that -- in the limit of optically dense winds -- the mass-loss scales in the same way as the terminal velocity for WR stars with the same stellar parameters at different metallicities. This is in sharp contrast to OB-star winds and an interesting testable prediction that could potentially provide an important observational mass-loss diagnostic. However, the emission-line dominated spectra will make it challenging to precisely pin down two stars with exactly the same stellar parameters.

Another important result from our modeling efforts is the metallicity-dependent breakdown of $\dot{M}$. As soon as the launching point of the wind approaches the optically thin regime, the linear trend is broken and the mass-loss rate rapidly decreases. At higher metallicity, more line opacities are available and thus a strong mass-loss can be maintained down to lower helium star masses. This rapid breakdown at different regimes qualitatively explains the empirical finding of different minimum luminosities for observed WN stars at different metallicities \cite[(see Shenar et al.\ 2020 for a more detailed discussion and implications on the role of the binary channel to form WR stars)]{Shenar+2020}. 

Incorporating both asymptotic behaviors of our model sequences, we find:

	\begin{equation}
	  \log \frac{\dot{M}}{M_\odot\,\mathrm{yr}^{-1}} =  a \cdot \log\left[-\log\left(1-\Gamma_\mathrm{e}\right)\right] - \log(2) \cdot \left(\frac{\Gamma_{\mathrm{e},\mathrm{b}}}{\Gamma_\mathrm{e}}\right)^c + \log \frac{\dot{M}_\mathrm{off}}{M_\odot\,\mathrm{yr}^{-1}}
  \end{equation}
	\begin{eqnarray}
	   \nonumber
	  \mathrm{with}\mbox{\hspace{1cm}}  a &=& 2.932 (\pm 0.016) \\
    \nonumber
		\Gamma_{\mathrm{e},\mathrm{b}} &=& -0.324 (\pm 0.011) \cdot \log(Z/Z_\odot) + 0.244 (\pm 0.010) \\
		\nonumber
		c &=& -0.44 (\pm 1.09) \cdot \log(Z/Z_\odot) + 9.15 (\pm 0.96) \\
		\nonumber
		\log \frac{\dot{M}_\mathrm{off}}{M_\odot\,\mathrm{yr}^{-1}} &=& 0.23 (\pm 0.04) \cdot \log(Z/Z_\odot) - 2.61 (\pm 0.03)
  \end{eqnarray}

This description marks the first theory-based mass-loss recipe for cWR stars. We do not vary the stellar temperature here and thus we do not expect that our results are valid for much cooler (late-type) WR stars. It is further worth mentioning that we do not account for different chemical abundances in this formula and thus quantitative differences are to be expected for stars with different composition. This is particularly important for stars with leftover hydrogen and for WC stars. However, we expect to see similar trends in follow-up studies for these groups. The results from our pilot study in \cite[Sander et al.\ (2020)]{Sander+2020} where we investigated both WC and WN models revealed a similarity in the trends for WN and WC models (both without hydrogen). However, for the same stellar parameters, the WC mass loss is expected to be lower than the WN mass loss.
This is a result of the smaller amount of Thomson opacity in a WC composition, which cannot be compensated by carbon line opacities as the ionization stages of carbon are too high in the deeper layers to notably contribute to the radiative driving. 
Instead, the carbon line opacities should lead to higher terminal velocities for the same stellar parameters. Our finding of comparably lower mass-loss rates for WC stars is in sharp contrast to the widely used description of \cite[Nugis \& Lamers (2000)]{NugisLamers2000} which predicts a higher mass-loss for WC stars as the carbon abundance contributes to their $Z$-term.

\section{Evolution of Wolf-Rayet stars with updated mass-loss treatment}

\begin{figure}[htb]
\begin{center}
   \includegraphics[width=0.235\textwidth]{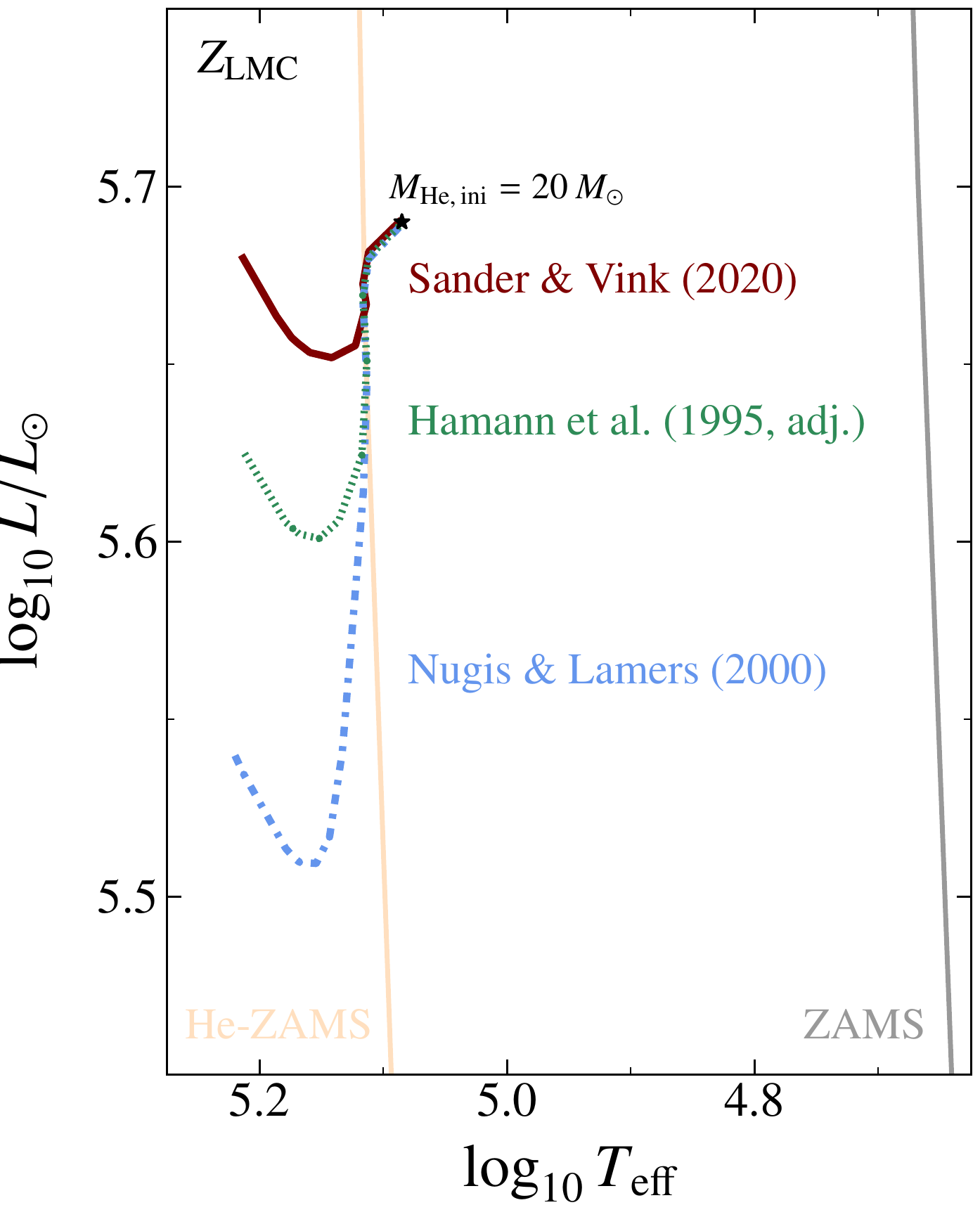} \hfill
   \includegraphics[width=0.38\textwidth]{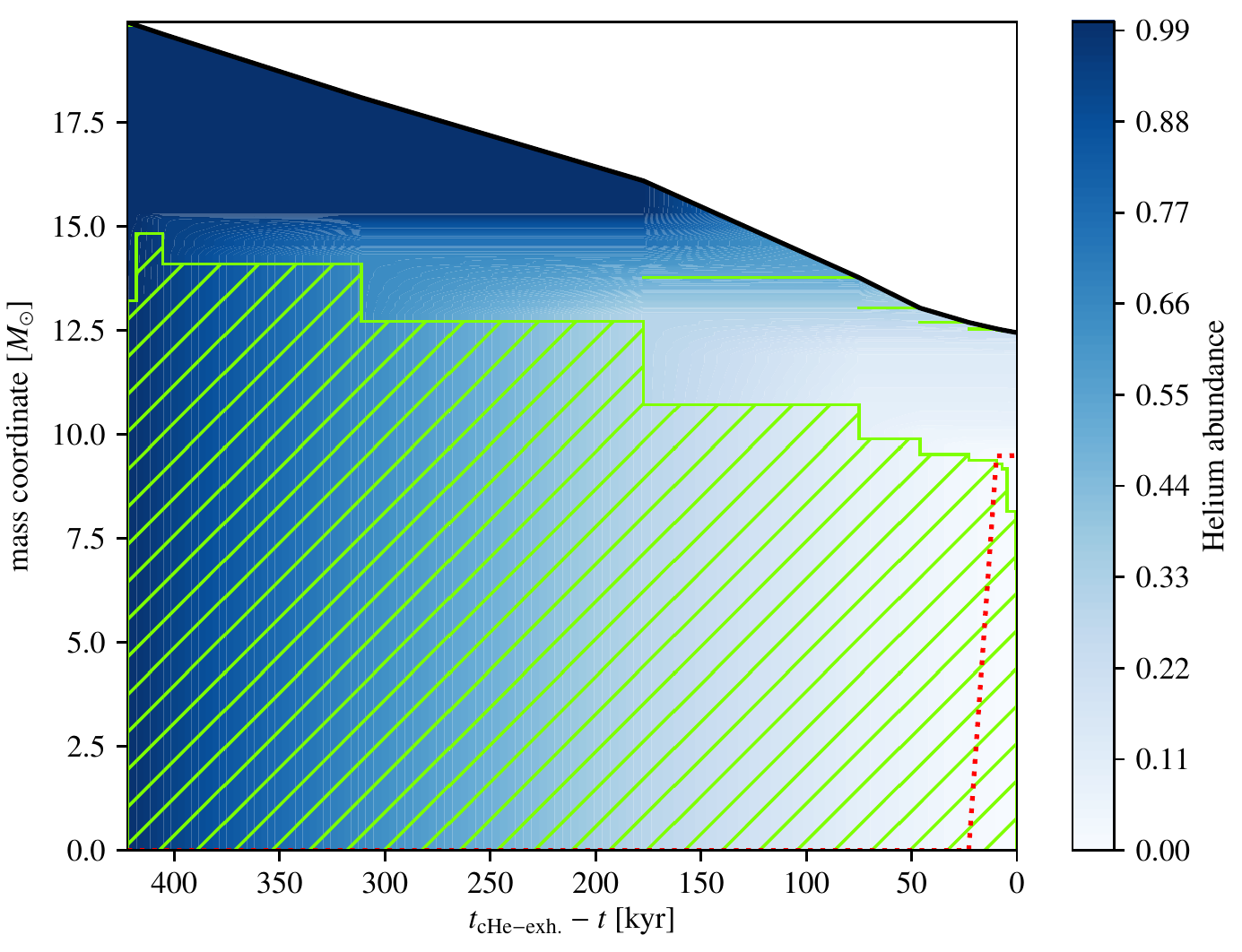} \hfill
 	 \includegraphics[width=0.295\textwidth]{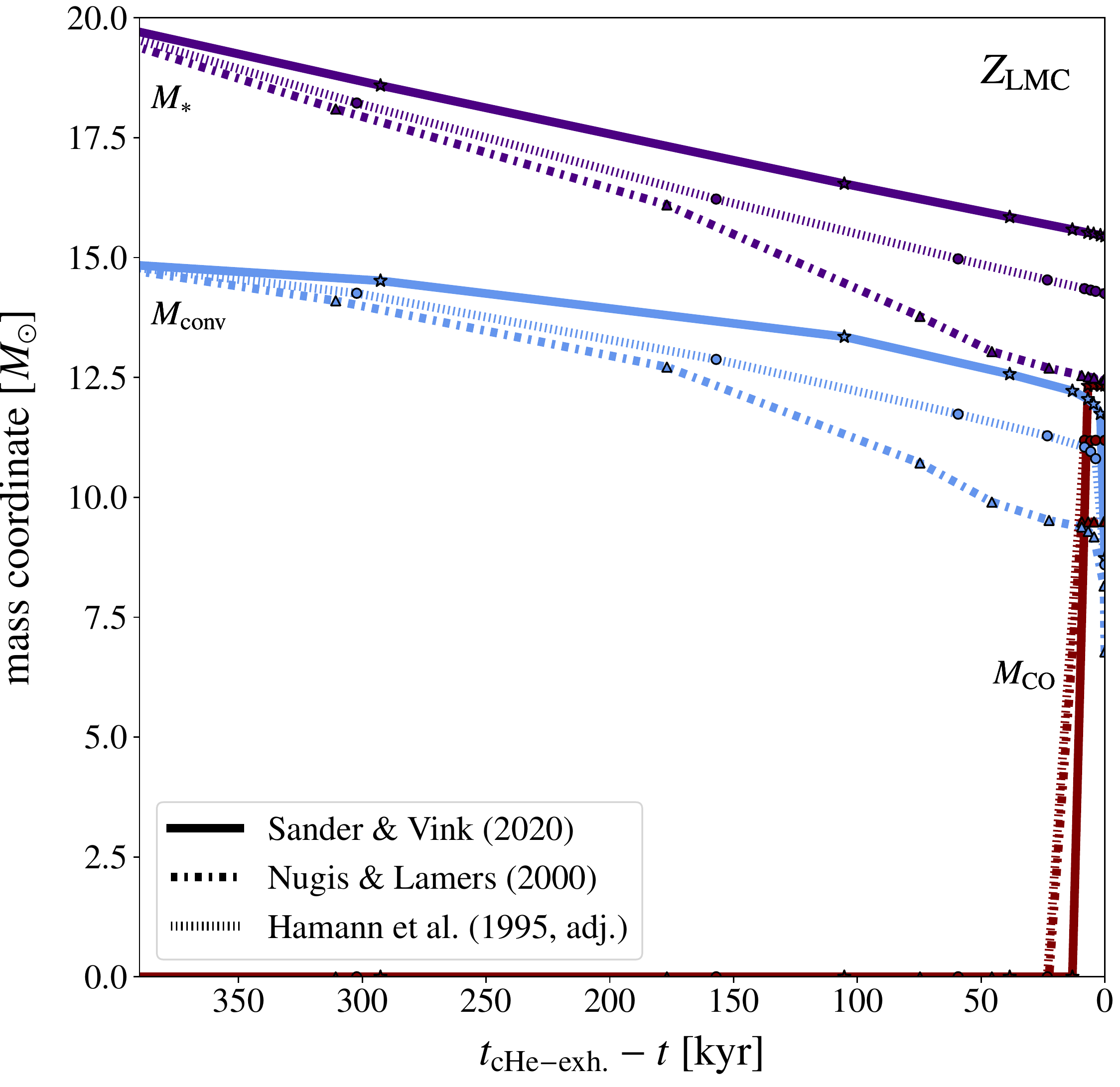}
 \caption{Left panel: Hertzsprung–Russell diagram for the 20\,$M_\odot$ He star models at $Z_\mathrm{LMC}$ using different $\dot{M}$ recipes. Middle panel: Kippenhahn diagram for the model using $\dot{M}$ from Nugis \& Lamers (2000). Right panel: Time evolution before core-He exhaustion of the total mass ($M_\ast$), convective core mass ($M_\mathrm{conv}$), and CO-core mass ($M_\mathrm{CO}$) for all three 20\,$M_\odot$ models at $Z_\mathrm{LMC}$.}
   \label{fig:wrevol}
\end{center}
\end{figure}

To investigate the impact of our improved mass-loss formalism we calculated a series of MESA helium star models \cite[(Higgins et al.\ 2021)]{Higgins+2021}. Each of the models was then evolved further with three different mass-loss treatments, namely our new formula from \cite[Sander \& Vink (2020)]{SanderVink2020}, the formula from \cite[Nugis \& Lamers (2000)]{NugisLamers2000}, and the treatment by \cite[Hamann et al.\ (1995)]{Hamann+1995} modified as suggested by \cite[Yoon et al. (2006)]{Yoon+2006}, i.e.\ divided by 10 and scaled with a metallicity-term according to \cite[Vink \& de Koter (2005)]{VinkdeKoter2005}. In our evolution models, metallicities between $Z_\odot$ and $0.02\,Z_\odot$ were covered with initial helium star masses ranging from $20\,M_\odot$ to $200\,M_\odot$. Since we were mainly interested deducing overall trends, we used the \cite[Sander \& Vink (2020)]{SanderVink2020} formula also for the WC stage and ignored the slight reduction of $\dot{M}$ predicted in \cite[Sander et al.\ (2020)]{Sander+2020}. We ran the models until the exhaustion of core-He burning and investigated the final masses as well as the CO-core masses (see also Fig.\,\ref{fig:wrevol}).

While at $Z = Z_\odot$, the change in final masses due to the different recipes is rather mild, the outcome changes drastically at lower metallicities. In particular the treatment of \cite[Nugis \& Lamers (2000)]{NugisLamers2000} turns out to be problematic: The incorporation of the carbon and oxygen abundances to the $Z$-term mimics a ``self-enrichment'' of the stars, thus leading to unrealistically high mass-loss rates. In fact, the usage of the recipe leads to an effective ``mass convergence'' that is almost independent of $Z$ (see also Fig.\,\ref{fig:wrtype-mdotdvinf}). This means that down to $0.1\,Z_\odot$ one would always end up with approximately the same final masses if a star manages to reach the cWR stage. This effect is avoided with the other two recipes, as only the initial metallicity $Z$ (or simply the iron abundance) enters here.

Another effect that is clearly visible in our evolution models is caused by the breakdown of WR-type winds predicted in \cite[Sander \& Vink (2020)]{SanderVink2020}: While the other two recipes lead to the loss of a few solar masses even at the lowest masses and metallicities, this is not the case in the new formula. This has notable consequences for the range of black hole masses that can be reached in the young Universe. Lower wind mass-loss rates yield higher black hole masses and thus could be an important part of the puzzle found by the black hole mass measurements from gravitational wave events. 

For higher-mass He stars, the reduced mass-loss at lower metallicity has another consequence: The CO core at the end of core-He burning can grow larger and thus the regime of pair instability could be reached more often. In fact, our results show that pair instability events could already occur below $0.5\,Z_\odot$ if sufficiently massive He stars can be formed. Previous mass-loss treatments would have shifted this limit down to $0.1\,Z_\odot$ or below, in particular when using \cite[Nugis \& Lamers (2000)]{NugisLamers2000}. All the details of our evolution models and a more in-depth discussion of the results can be found in \cite[Higgins et al.\ (2021)]{Higgins+2021}. 

\section{Summary \& Conclusions}

Using sequences of dynamically-consistent atmosphere models, we derived the very first theory-based mass-loss description for classical Wolf-Rayet stars \cite[(Sander \& Vink 2020)]{SanderVink2020}. Our models represent stars with a WN-type composition. The sequences reveal a metallicity-dependent breakdown of WR-type mass loss when transiting to thinner winds. In the limit of optically thick winds, we derive a very modest metallicity scaling of $\dot{M} \propto Z^{0.3}$. Moreover, the ratio of mass-loss rate over the terminal velocity seems to be constant in this limit, independent of metallicity. We propose that the empirical relations which predict stronger metallicity-scalings are a consequence of the beginning breakdown of WR-type mass loss, which leads to a deviation from the rather flat $\dot{M}$-slope in the optically thick limit. The derived velocity fields are complex and deviate significantly from simple $\beta$-law descriptions. This is a consequence of the multiple ionization changes in a WR-type wind with the inner wind being launched due to iron M-shell opacities.

To reach the regime of WR-type mass loss at lower metallicities and thus lower iron abundances, cWR stars have to get closer to the Eddington limit, similar to what \cite[Gr{\"a}fener \& Hamann (2008)]{GraefenerHamann2008} found for WNh-type stars. This is in line with empirical findings that the minimum luminosity for WR stars becomes higher at lower metallicity \cite[(Shenar et al.\ 2020)]{Shenar+2020}. Below the breakdown of WR-type mass loss, our models are also huge sources of He\,\textsc{ii} ionizing flux. This is an immediate consequence of the weaker winds as the radiation in the extreme UV is otherwise used to drive the WR winds and only re-emitted at much longer wavelengths.  Using a series of MESA models for helium stars, we demonstrate that implementing the different scaling of WR-type mass loss and its breakdown is crucial for deriving proper black hole masses. The lack of WR-type mass loss at lower metallicity makes it not only easier to form massive black holes, but also allows to reach the regime of pair instability up to $\approx 0.5\,Z_\odot$ as long as massive enough He stars can be formed.

Our results presented in these proceedings and the cited papers mark only the first efforts on the way to a full comprehension of WR-type mass loss. Beside expanding to a broader set of stellar parameters and further refinements to account for different chemical compositions, the absolute values of $\dot{M}$ will also depend on a better description of multi-dimensional effects such as a more realistic treatment of clumping and its still quite enigmatic onset in WR winds.

\end{document}